\documentstyle[12pt,aaspp4]{article}
\begin{document}

\title{Extracting Energy from Accretion into Kerr Black Hole}

\author{Li-Xin Li and Bohdan Paczy\'nski}
\affil{Princeton University Observatory, Princeton, NJ 08544--1001, USA}
\affil{E-mail: lxl, bp@astro.princeton.edu}

\begin{abstract}
The highest efficiency of converting rest mass into energy by accreting
matter into a Kerr black hole is $ \sim 31\% $ (Thorne 1974).  
We propose a new
process in which periods of accretion from a thin disk, and the associated
spin-up of the black hole, alternate with the periods of no accretion and
magnetic transfer of energy from the black hole to the disk.  These cycles
can repeat indefinitely, at least in principle, with the black hole
mass increasing by $ \sim 66\% $ per cycle, and up to $ \sim 43\% $
of accreted rest mass radiated away by the disk.
\end{abstract}

\keywords{black hole physics --- accretion disks --- magnetic fields}

\section{Introduction}
Accretion of matter from a thin disk into a black hole may spin up
the black hole up to $ a^{\star} = 0.998 $, at which point $ \sim 31\% $
of the rest mass of accreted matter is radiated away (Thorne 1974).
We define $ a^{\star} \equiv c a / G M_H $, where $ M_H $ is
the black hole mass and $ a $ is its angular momentum per unit mass.
Energy may be extracted out of a rapidly spinning black hole
by means of the Blandford-Znajek mechanism and transported
to a distant load by means of magnetic field (Blandford \&
Znajek 1977).  Recently it
has been shown that even more efficient extraction of rotational
energy is possible with the magnetic field lines threading the Kerr
black hole and connecting it to the inner parts of the disk (Li 2000).
In these two processes energy of black hole rotation may be extracted
with up to 9\% and 15\% efficiency, respectively (Li 2000).

In this letter we propose an efficient way of converting rest mass
to energy by alternating the two processes: accretion of matter from a
thin disk into a Kerr black hole and the magnetic extraction of
energy from the black hole to the disk. We show that this process
has the highest efficiency in converting mass into energy through
accretion into a black hole among those have been ever suggested:
up to $ \sim 43\% $ of accreted rest mass is radiated away by the disk,
with the black hole mass increasing by $ \sim 66\% $ per cycle.

\section{Results}
We consider a black
hole with the initial mass $M_{A1}$ and spin
$a^{\star}_{A1}$.  Next, the black hole is spun up by matter accreting
from a thin, Keplerian disk; its mass and spin
increase to $M_{B1}$ and $a^{\star}_{B1}$, respectively.
During this phase the amount of accreted
disk rest mass is $M_{d1}$, and the total energy $E_{A1}$ is radiated
by the disk.  Next the accretion stops, and the black hole
transfers angular momentum and energy to the disk by means of magnetic
field, as described by Li (2000), reducing the mass and the spin to the
new values $M_{A2}$ and $a^{\star}_{A2}$, and transferring energy
$E_{B1}$ to the disk.  That energy is radiated by the disk, while
no new matter is accreted into the black hole.

During the evolution from $ A_1 $ to $ B_1 $ we calculate the change
of black hole mass and spin, as well as the efficiency of converting
accreted mass to radiation, using equations [3] and [4] provided by
Bardeen (1970).  In this calculation we have used Bardeen's simple
solutions for accretion and the effects of `photon capture' of
Thorne (1974) have not been taken into account, except that $ a^{\star} $
has been limited to values not exceeding $ a^{\star}_{max} = 0.998 $.
Note that the efficiency
of converting rest mass to radiation by accretion from a thin disk into
$ a^{\star} = 0.998 $ Kerr black hole is equal to 32.4\% in
Bardeen's formulation, but it is only 30.8\% in the Thorne's model.
However, this difference becomes insignificant when the $ a^{\star} $
parameter is somewhat smaller than 0.998.  Therefore, we overestimate
the efficiency of energy release in during the evolution from $ A_1 $
to $ B_1 $ by a small amount only.
During the evolution from $ B_1 $ to $ A_2 $ we use equation [13]
as given by Li (2000).

We consider a cycle which could be repeated indefinitely, and therefore
we seek $ a^{\star}_{A1} = a^{\star}_{A2} $.  We would like
the process to be as efficient as possible in converting the accreted disk
mass $ M_{d1} $ to the radiated energy $ E_1=E_{A1}+A_{B1} $.  Hence,
we seek the values of $ a^{\star}_{A1} $ and $ a^{\star}_{B1}$ which maximize
the ratio $ E_1/M_{d1} c^2 $.
It turns out that the optimum values are:
$ a^{\star}_{A1} = a^{\star}_{A2} \approx 0.3594 $, and
$ a^{\star}_{B1} \approx 0.998 $,
which give the efficiency of converting disk rest
mass to radiation: $ \eta _{max} = E_1/M_{d1} c^2 \approx 0.436 $.

The evolution of our system is shown in Fig. 1:
the black hole mass $M_H$ varies as a function of integrated
amount of energy radiated from the disk $E$.
Obviously, $ E $ varies monotonically with time,
but $ M_H $ increases during the accretion phase (from $A_1$ to $B_1$),
and it is reduced during the spin-down phase (from $B_1$ to $A_2$).
The black hole mass increases
by the same factor in every cycle, and it can grow as long as the
cycles continue.

During the first part of
the cycle the black hole mass increases by a factor 1.9625 between
points $ A_1 $ and $ B_1 $, while the amount of rest mass accreted from
a thin disk is equal to $ 1.1788 M_{A1} $, where $ M_{A1} $ is the initial
black hole mass.  The balance, equal to $ 0.2163 M_{A1} c^2 $ is
radiated away.  In the second part of the cycle no mass is accreted,
the black hole mass is reduced to 1.6644 of its initial value, and
the corresponding amount of energy, i.e. $ 0.2981 M_{A1}c^2 $
is transferred to the disk and radiated away.  During the
complete cycle the black hole mass increases by a factor 1.6644, and
the total energy radiated from the disk is equal to $ E_1 = 0.5144 M_{A1} c^2
= 0.436 M_{d1}c^2 $, where $ M_{d1} $ is the total accreted disk mass.

This process can be repeated indefinitely, at least in principle.
In each cycle the black hole mass increases by a factor 1.6644.
The system evolves through alternating phases of disk accretion
associated with the black hole spin-up, and phases of no accretion
and the disk radiating away energy transferred from the spinning-down
black hole by means of the magnetic field.  The evolution of the system
through two full cycles is illustrated in Fig. 1.

\section{Conclusions}
It is far from clear if the scenario we propose can be maintained
over many cycles, or even over one cycle.  However, it shows that
it may be possible, at least in principle, to radiate up to $ \sim 43\% $
of rest mass energy, with only $ \sim 57\% $ ending up permanently in the
black hole.  This is the highest efficiency of converting rest mass to
energy by accreting into a black hole that has been ever proposed.

\acknowledgments{This work was partly supported by the NASA grant
NAG5-7016.}


\newpage
\figcaption[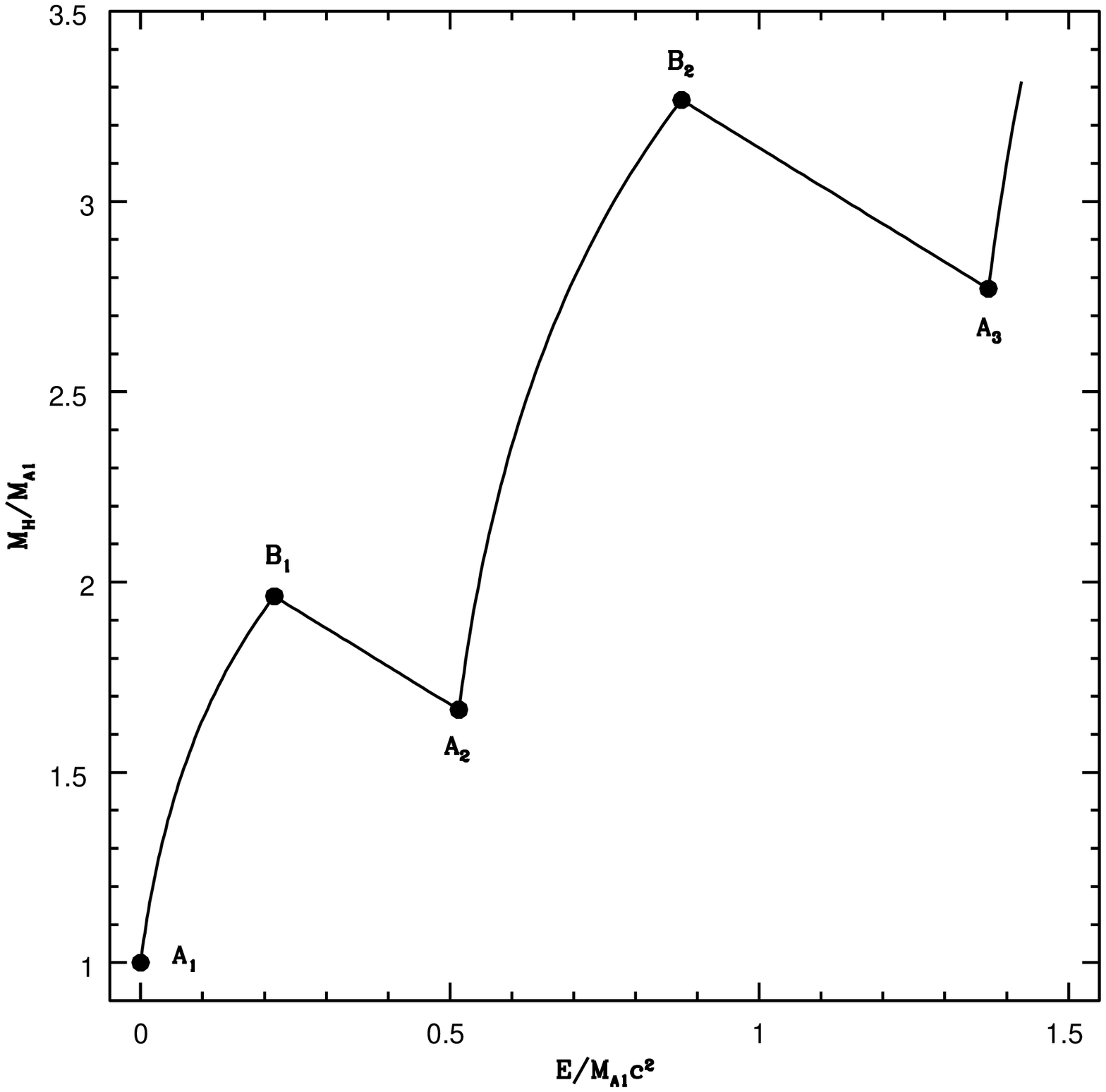]{The evolution of the black hole - disk system
is shown through two full cycles of the accretion - spin-down process,
with time increasing from left to right. The
black hole mass $ M_H $ in units of the initial mass $ M_{A1} $
is shown as a function of the integrated amount of energy radiated by
the disk, in units of $ M_{A1} c^2 $. Each cycle contains two phases:
phase $ A \rightarrow B $, during which the black hole mass and
spin increase due to accretion from a thin disk, and phase
$ B \rightarrow A $, during which there is no accretion and the
black hole transfers energy and angular momentum to the disk.
In every cycle the black hole mass increases by a factor $ \sim 1.66 $,
and $ \sim 43\% $ of the rest mass of accreted matter is radiated away.
\label{figure1}}

\end{document}